\documentclass{iau} 
\usepackage{graphicx,hyperref}

\title[The next-generation VLA] 
{A next-generation Very Large Array}

\author[Eric J. Murphy]   
{Eric J. Murphy\\ (on behalf of the ngVLA community)}

\affiliation{National Radio Astronomy Observatory, 520 Edgemont Road, Charlottesville, VA 22903, USA \\ email: {\tt emurphy@nrao.edu}}

\pubyear{2017}
\volume{336}  
\jname{Astrophysical Masers: Unlocking the Mysteries of the Universe}
\editors{A. Tarchi, M.J. Reid \& P. Castangia, eds.}
\begin{document}

\maketitle

\begin{abstract}
In this proceeding, we summarize the key science goals and reference design for a next-generation Very Large Array (ngVLA) that is envisaged to operate in the 2030s.  
The ngVLA is an interferometric array with more than 10 times the sensitivity and spatial resolution of the current VLA and ALMA, that will operate at frequencies spanning $\sim$1.2 -- 116\,GHz, thus lending itself to be highly complementary to ALMA and the SKA1.  
As such, the ngVLA will tackle a broad range of outstanding questions in modern astronomy by simultaneously delivering the capability to: unveil the formation of Solar System analogues; probe the initial conditions for planetary systems and life with astrochemistry; characterize the assembly, structure, and evolution of galaxies from the first billion years to the present; use pulsars in the Galactic center as fundamental tests of gravity; and understand the formation and evolution of stellar and supermassive blackholes in the era of multi-messenger astronomy.

\end{abstract}

\firstsection 
\section{Introduction}
\begin{figure}[!t]
\begin{center}
 \includegraphics[width=5in]{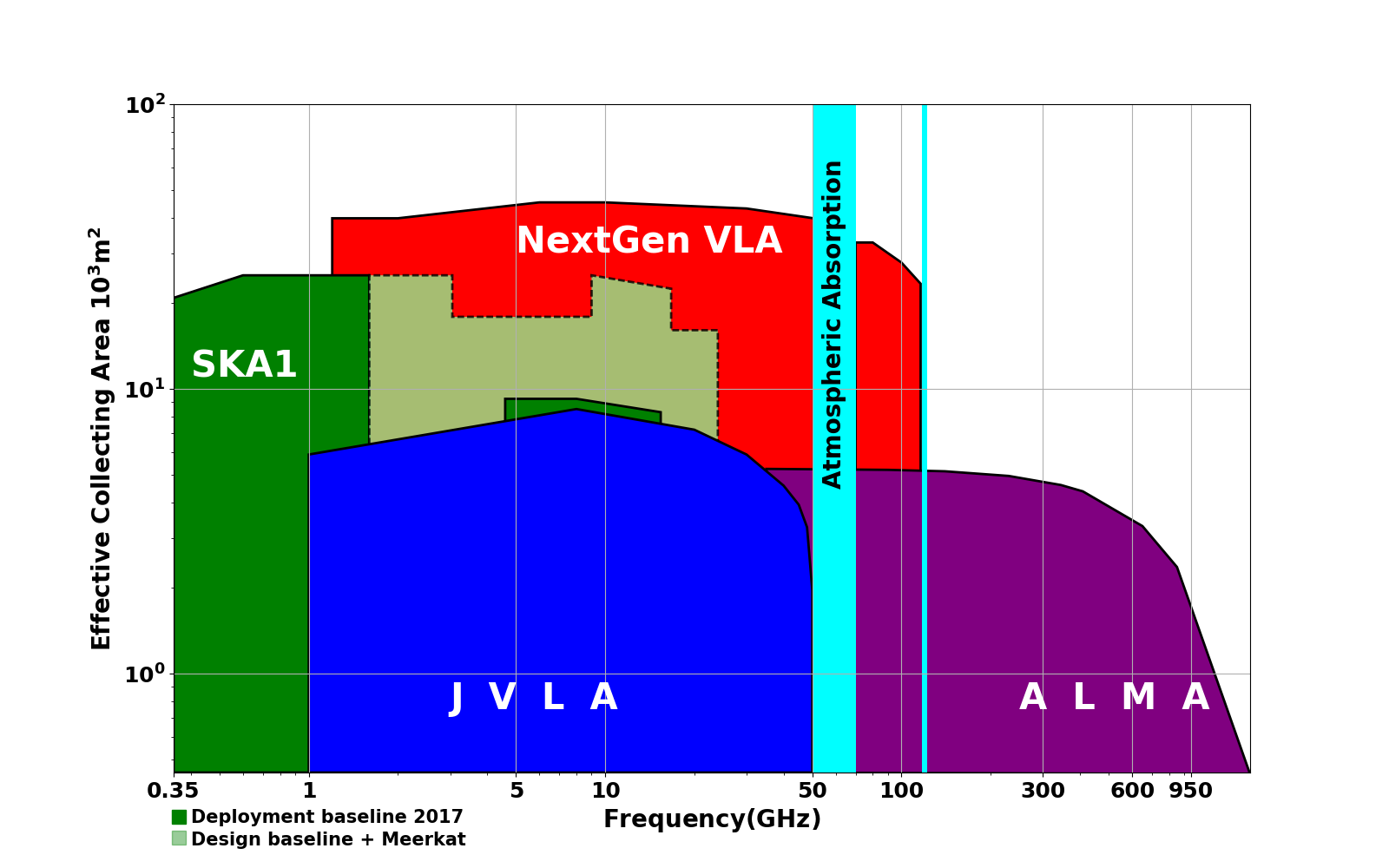} 
 \caption{ A comparison of effective collecting area plotted against frequency for various radio/mm facilities that will be operating in the 2030s, illustrating how complementary the ngVLA is to both ALMA and the SKA1.}
   \label{fig1}
\end{center}
\end{figure}

Inspired by dramatic discoveries from the Jansky VLA and ALMA, a plan to pursue a large collecting area radio interferometer that will open new discovery space from protoplanetary disks to distant galaxies is being developed by the astronomical community and NRAO.  
Building on the superb cm observing conditions and existing infrastructure of the VLA site, the current vision of a next-generation Very Large Array (ngVLA) will be an interferometric array with more than 10 times the sensitivity area and spatial resolution of the current VLA and ALMA, that will operate at frequencies spanning $\sim$1.2 -- 116\,GHz.  
The ngVLA will be optimized for observations at wavelengths between the exquisite performance of ALMA in the submm, and the future SKA1 at decimeter to meter wavelengths, thus lending itself to be highly complementary with these facilities.  
This is illustrated in Figure \ref{fig1} where the effective collecting area for a number of radio/mm facilities expecting to be operational in the 2030s is shown.  
The ngVLA clearly sits in the frequency range outside of the core science missions for both the SKA1 and ALMA, creating an overwhelming set of new synergistic opportunities with these other facilities.   

As such, the ngVLA will open a new window on the universe through ultra-sensitive imaging of thermal line and continuum emission down to milliarcecond resolution, as well as deliver unprecedented broad band continuum polarimetric imaging of non-thermal processes yielding a broad range of scientific discoveries (e.g., planet formation, signatures of pre-biotic molecules, cosmic cycling of cool gas in galaxies, massive star formation in the Galaxy etc.)
Additionally included in current ngVLA planning is an option to greatly expand current U.S. VLBI capabilities by both replacing existing VLBA antennas/infrastructure with ngVLA technology and adding additional stations on $\sim$1000\,km baselines to bridge the gap between ngVLA and existing VLBA baselines.  
A second science option to provide access to the low frequency sky (i.e., $5-800$\,MHz) in a commensal fashion is also being explored.  

\section{ngVLA Key Science Goals}

The ngVLA Science Advisory Council (SAC), a group of experts appointed by NRAO, in collaboration with the broader international astronomical community, recently developed over 80 compelling science cases requiring observations between 1.2 -- 116\,GHz with sensitivity, angular resolution, and mapping capabilities far beyond those provided by the Jansky VLA, ALMA, and the SKA1. 
These science cases span a broad range of topics in the fields of planetary science, Galactic and extragalactic astronomy, as well as fundamental physics.
Consequently, the primary science requirement for the ngVLA has overwhelmingly been determined to be flexible enough to support the wide breadth of scientific investigations that will be proposed by its highly creative user base over the full lifetime of the instrument. 
This in turn makes the ngVLA a different style of instrument than many other facilities on the horizon (e.g., SKA1, LSST, etc.), which are heavily focused on carrying out large surveys.  
However, each of the individual science cases were objectively reviewed and thoroughly discussed by the different Science Working Groups within the ngVLA-SAC, ultimately distilling a finite list of key scientific goals for a future radio/mm telescope. 
The initial set of key science goals, along with the results from the entire list of science use cases, were then presented and discussed with the broader community at the ngVLA Science and Technology Workshop June 26 -- 29, 2017 in Socorro, NM as a means to build consensus around a single vision for the key science mission of the ngVLA\footnote{\url{https://science.nrao.edu/science/meetings/2017/ngvla-science-program/index}}.  
These key science goals are described in detail in \cite{memo19} and include:
\begin{itemize}
\vspace*{6 pt}
\item Unveiling the Formation of Solar System Analogs
\vspace*{2 pt}
\item Probing the Initial Conditions for Planetary Systems and Life with Astrochemistry
\vspace*{2 pt}
\item Charting the Assembly, Structure, and Evolution of Galaxies from the First Billion Years to the Present
\vspace*{2 pt}
\item Using Pulsars in the Galactic Center to Make a Fundamental Test of Gravity
\vspace*{2 pt}
\item Understanding the Formation and Evolution of Stellar and Supermassive Black Holes in the Era of Multi-Messenger Astronomy 
\vspace*{6 pt}
\end{itemize}

\begin{figure}[!t]
\begin{center}
 \includegraphics[width=4in]{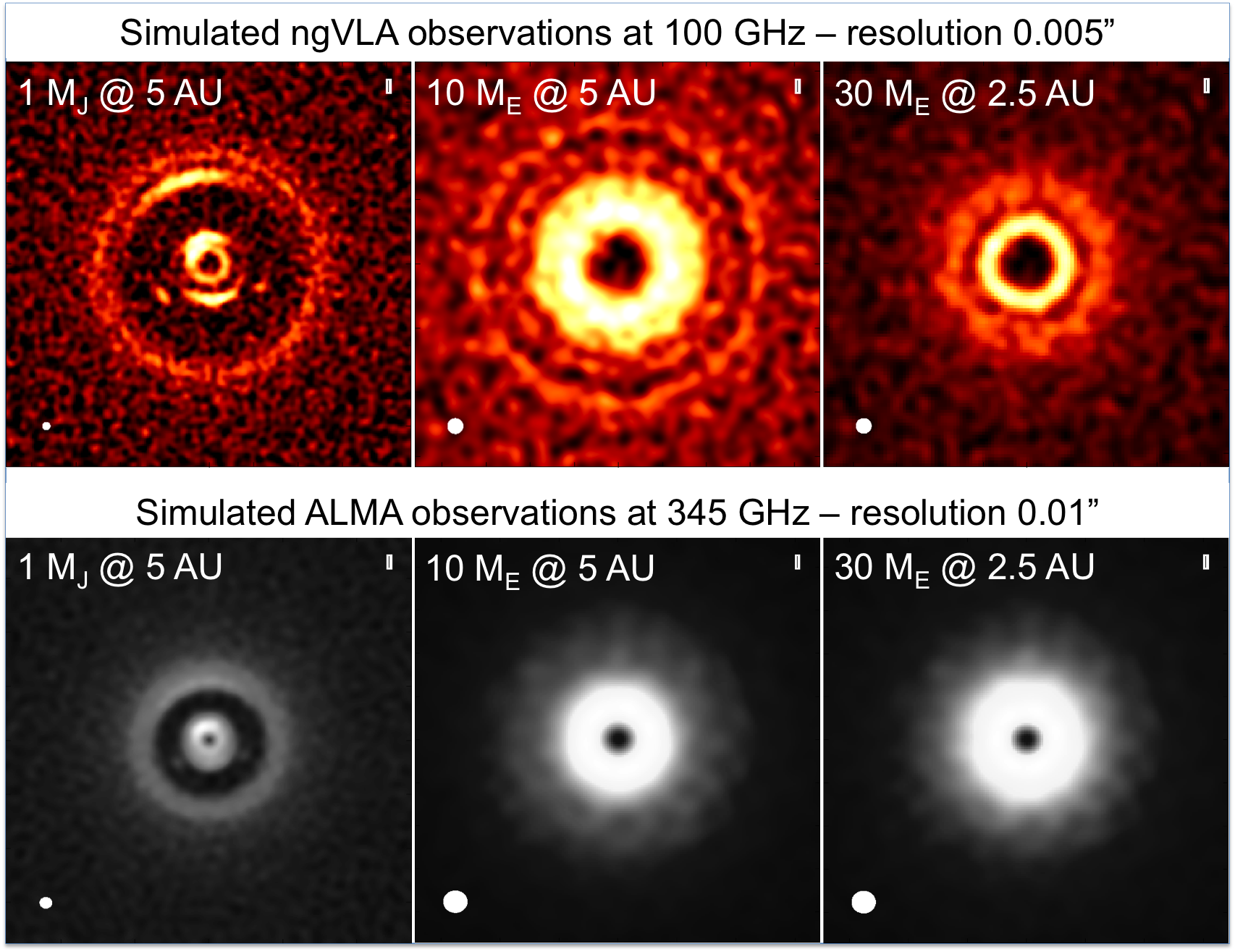} 
 \caption{ Simulated observations (ngVLA on top row, ALMA on bottom row) of the continuum emission of a protoplanetary disk perturbed by a Jupiter mass planet orbiting at 5 au (left), a 10 Earth mass planet orbiting at 5 au (center), and a 30 Earth-mass planet orbiting at 2.5 au (right). The ngVLA observations at 100\,GHz were simulated assuming an angular resolution of 5\,mas and a rms noise level of 0.5 $\mu$Jy beam$^{-1}$. ALMA observations at 345 GHz where simulated assuming the most extended array configuration comprising baselines up to 16\,km and a rms noise level of 8\,$\mu$Jy beam$^{-1}$. From \cite{ricci18}.
 }
   \label{fig2}
\end{center}
\end{figure}

In Figure \ref{fig2} we highlight the power of the ngVLA to execute one of these science goals.  
The figure shows a comparison of ngVLA- (top row) and ALMA- (bottom row) simulated observations of continuum emission of a protoplanetary disk perturbed by a Jupiter mass planet orbiting at 5 au (left column), a 10 Earth mass planet orbiting at 5 au (center column), and a 30 Earth mass planet orbiting at 2.5 au (right column).  
ALMA is only able to achieve au-scale resolution at $\gtrsim$345\,GHz, where the associated emission from such disks is optically-thick.  
The ngVLA will deliver the requisite combination of frequency coverage, resolution, and sensitivity to pierce into these highly enshrouded regions and directly image planet formation in the terrestrial zone.  
Even more spectacular, by imaging such sources with a monthly cadence over many years, the ngVLA effectively turns planet formation into a time-domain science, by being able to track the planetary orbits and associated tidal debris and see how they evolve.  
Such observations are highly synergistic with the major goals of both ground-based 30\,m glass optical, and currently discussed next-generation UV/optical/NIR space missions, which include direct imaging of planets in the terrestrial zone.  

\begin{figure}[!t]
\begin{center}
 \includegraphics[width=5in]{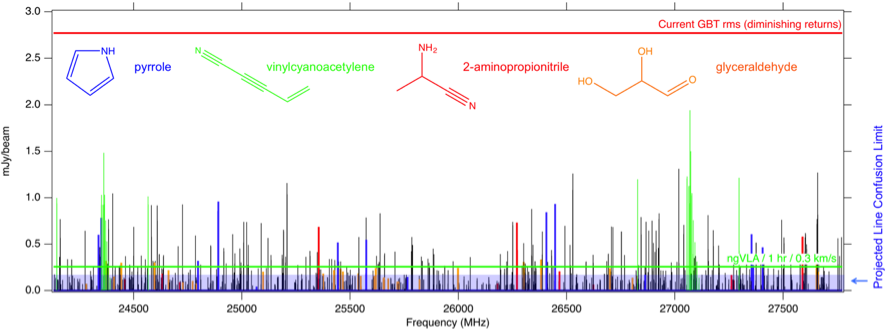} 
 \caption{ A conservative simulation of 30 complex organic molecules not currently detected in the ISM (in black), but which are good candidates for detection by the ngVLA.  
 If present, such species are expected to be detectable by the ngVLA above the confusion limit of a typical ngVLA line-survey.  
 A few key molecules are highlighted in color, which are important for understanding chemical evolution (e.g., increasingly complex carbon-nitrogen chains), probing chirality (e.g., glyceraldehyde), are part of largely unknown cyclic/aromatic chemistry (e.g., pyrrole), or are excellent candidates for being direct precursors to amino acids and other biogenic species (e.g., aminopropionitrile).  (Credit: B. McGuire)
 }
   \label{fig3}
\end{center}
\end{figure}

The ngVLA will also be able to uniquely access and characterize a large range of predicted, yet undetected, pre-biotic molecules to try to discern the necessary requirements for habitability in proto-stellar regions.  
This is a second key science goal of the ngVLA.  
A simulated spectrum of 30 such complex organic molecules is shown in Figure \ref{fig3} (Credit: B. McGuire).  
Each of these molecules are currently undetected in the ISM and clearly out of reach using current facilities.   
However, if present, such species are expected to be detectable by the ngVLA at frequencies where line blending is not an issue.  
A few key molecules are highlighted in color (see Figure \ref{fig3} caption).
The simulation assumes that the emission is coming from a compact, 1$^{\prime\prime}$~hot core in Sgr B2(N), from molecules with column densities of $10^{12} - 10^{14}$\,cm$^{-2}$, at $T = 200$\,K, and with a linewidth of 3\,km\,s$^{-1}$.

While we have highlighted two of the key ngVLA science goals here, as stated above, the primary mission of the ngVLA is to be flexible enough to deliver capabilities, enabling a broad range of science.    
A brief list includes: 
gravitational wave EM follow-up, 
extrasolar space weather, 
studying the bursting universe (FRB, GRB, TDE, etc.), 
low surface brightness H{\sc i} and  CO, 
obscured black hole growth and associated AGN physics, 
quasar-mode feedback and the SZ effect, 
black hole masses and H$_{0}$ with mega-masers, 
$\mu$as Astrometry: ICRF, Galactic structure, etc., 
solar system remote sensing: passive and active radar, 
spacecraft telemetry, tracking: movies from Mars, and much more.

\section{Reference Design Summary}

The ngVLA concept began to rapidly converge post the June 2017 workshop in both its requirements and a matching concept that was able to support more than 80\% of all of the science use cases submitted by the community (see \cite[Selina, Murphy, \& Erickson 2017]{memo18}).  
A more detailed description of the design and the arrays performance can be found in \cite{memo17}.  
Here, we simply highlight a number of the key elements.  

\begin{figure}[!t]
\begin{center}
 \includegraphics[width=3.4in]{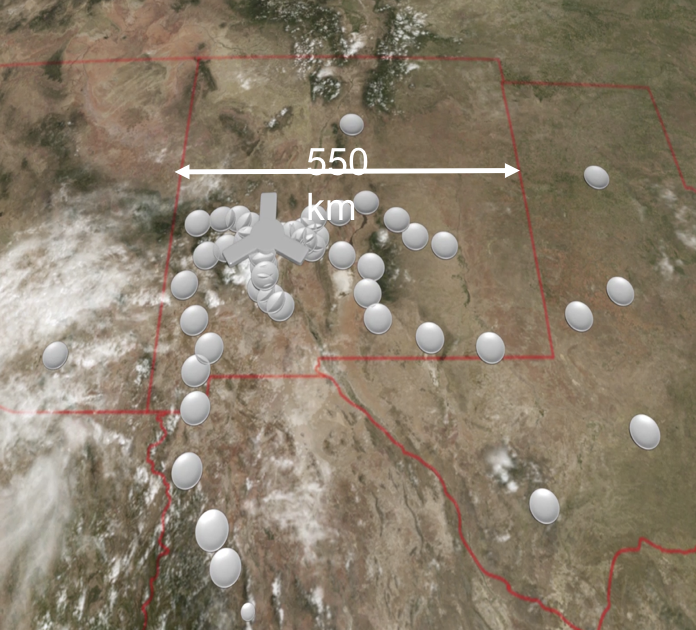} 
 \caption{ Stations in the present ngVLA configuration (Greisen \& Owen). 
 The compact core is located at the apex of the present VLA, and antennas are populated along the VLA arms. 
 Long baseline stations radiate primarily south and east from the VLA. The locations of the long baseline stations are approximate, but account for land ownership and available infrastructure including roads, electrical distribution lines, and fiber optics networks.}
   \label{fig4}
\end{center}
\end{figure}

The array configuration (Figure \ref{fig4}) shows a notional picture of the extent of the dish locations having a longest baselines of nearly 1000\,km.  
The array collecting area is distributed to provide high surface brightness sensitivity on a range of angular scales spanning from approximately 1000 to 5\,mas. 
In practice, this means a core with a large fraction ($\approx$50\%) of the collecting area in a randomized distribution to provide high snapshot imaging fidelity, and arms/rings extending asymmetrically out to $\approx$1000\,km baselines, filling out the $(u,v)$-plane with Earth rotation and frequency synthesis.  
The exact configuration remains a work in progress, as we try to identify a version that yields the highest quality synthesized beam for a large range of angular resolutions.  

The present concept is a homogeneous array of 214 18m apertures, which is supported by an internal parametric cost estimation.   
The antenna locations are fixed, and each is outfitted with front ends that provide access to the atmospheric windows spanning 1.2 -- 50.5\,GHz and 70 -- 116\,GHz.  
The current front-end concept uses 6 single-pixel feeds in two dewar packages, but a single dewar solution is actively being pursued.  
The antenna optical configuration favors unblocked apertures, with an Offset Gregorian feed-low design offering synergy with the front-end concepts under consideration, while additionally accounting for maintenance and operational concerns. The antenna surface error will be limited to of order 160\,$\mu$m rms, ensuring that the antenna Ruze efficiency is better than 50\% at the 116\,GHz upper operating limit.
The correlator is an FX design, possibly with a distributed F-engine. 
The correlator will support up to 64\,k channels at coarse time resolution, or up to 1\,msec time resolution for time-domain science. 
The design will incorporate other necessary observing modes such as phased arrays for VLBI and pulsar timing or searches. 

While there has been significant convergence for many of the design parameters, there still remains a number of open questions in the reference design that are actively being investigated.  
These include determining the best strategy for phase calibration, an optimized array configuration that delivers a high-quality synthesized beam for a large range of angular scales, and the need and associated requirements for a short-spacing\,mas array (e.g., 16 6\,m antenna array combined with 18\,m total power antennas) to recover missing flux.


\section{Additional Science Options}
\subsection{Expansion of VLBI Capabilities}
We are additionally looking at a possible science option that will upgrade/replace existing VLBA antennas/infrastructure with ngVLA technology.  
Such an option would be outside of the ngVLA project (construction and operations), and may largely depend on resources external from the astronomy community.  

A potential configuration for such an option is to have 10 -- 20\% of the ngVLA collecting area be placed in $\approx$8 stations of 3 antennas each.  
This will introduce new $\sim$1000\,km baseline stations that will bridge the gap between the ngVLA core's ``short" (50 -- 500\,km) spacings and the current continental-scale VLBA baselines.  
New stations, and any VLBA antenna replacements would leverage the ngVLA antenna design and receiver package, with all data being electronically transferred using the full ngVLA bandwidth.  

Such a facility would significantly improve upon the current VLBA capabilities (sensitivity and astrometric accuracy), yielding a number of exciting new scientific possibilities.  
These include, sub- 1\% constraints on H$_{0}$ measurements using megamasers, an order of magnitude improvement in General Relativity parameters from light bending, and routine distance measurements across the Milky Way.    
Such capabilities may even open the door to imaging the surfaces of 10's to 100's of stars. which is highly synergistic with current optical interferometry goals.  

\subsection{Commensal Low Frequency Science}
There is also an investigation looking into the possibility for commensal low frequency capabilities, which is currently being called the next-generation LOw Band Observatory (ngLOBO; see \cite[Taylor et al. 2017]{memo20}).  
ngLOBO would provide access to the low frequency sky (i.e., $5-800$\,MHz) in a commensal fashion, operating independently from the ngVLA, but leveraging common infrastructure (e.g., land, fiber, power, etc.) . 
This approach provides continuous coverage through an aperture array (called ngLOBO-Low) below 150\,MHz and by accessing the primary focus of the ngVLA antennas (called ngLOBO-High) above 150\,MHz. 

ngLOBO has three primary scientific missions: 
(1) Radio Large Synoptic Survey Telescope (Radio-LSST): a commensal, continuous synoptic survey of large swaths of the sky for both slow and fast transients; 
(2) Complementary low frequency images of all ngVLA targets and their environments to enhance their value;  
(3) Independent beams from the ngLOBO-Low aperture array for research in astrophysics, Earth science and space weather applications, engaging new communities and attracting independent resources. 
The ngVLA will be a superb, high frequency instrument and ngLOBO will additionally allow it to participate in the worldwide renaissance of low frequency science.

\section{Summary}
The ngVLA is being designed to tap into the astronomical community's intellectual curiosity and enable a broad range of scientific discoveries (e.g., planet formation, signatures of pre-biotic molecules, cosmic cycling of cool gas in galaxies, massive star formation in the Galaxy etc.).  
Based on community input to date, the ngVLA is the obvious next step to build on the VLA's legacy and continue the U.S.'s place as a world leader in radio astronomy.  
Presently, there have been no major technological risks identified.  
However, the project is continually looking to take advantage of major engineering advancements seeking performance and operations optimizations.  
As the project moves forward in preparation for the U.S. 2020 Astronomy Decadal Review, we will continue to refine the ngVLA science mission and instrument specifications/performance through a detailed science book and reference design study.  
The ultimate goal of the ngVLA is to give the U.S. and international communities a highly capable and flexible instrument to pursue their science in critical, yet complementary ways, with the large range of multi-wavelength facilities that are on a similar horizon.

\end{document}